\documentclass[prc,tightenlines,a4paper,nofootinbib]{revtex4}
\usepackage{amsmath,amssymb,graphicx}
\usepackage{bm}% bold math
\newcommand{\la}{\langle}
\newcommand{\ra}{\rangle}
\newcommand{\be}{\begin{equation}}
\newcommand{\ee}{\end{equation}}
\newcommand{\nn}{\nonumber\\}

\begin{document}
\title{Renormalisation-group analysis of electromagnetic couplings\\ in the 
pionless effective field theory}

\author{A. N. Kvinikhidze}

\affiliation{Department of Theoretical Physics, A. Razmadze Mathematical Institute, I. Javakhishvili Tbilisi State University, GE-0177 Tbilisi, Georgia}

\author{M. C. Birse}

\affiliation{Theoretical Physics Division, School of Physics and Astronomy,
The University of Manchester, Manchester, M13 9PL, UK}

\begin{abstract}

The Wilsonian renormalisation group is applied to a system of two nonrelativistic particles interacting via short-range forces and coupled to an external EM field. By demanding that a fully off-shell one-particle-irreducible 5-point amplitude is independent of the cutoff, a renormalisation group equation is derived for the interaction current density. This is solved to obtain the fixed point corresponding to the unitary limit. The scaling behaviour of perturbations around this point is analysed. Some of these terms are related by gauge invariance to terms in the effective-range expansion; others describe short-range physics that is not included in the two-body potential. We construct observables including the bound-state form factor and show how the scaling of the terms in the interaction current is reflected in the power counting for their contributions to observables. 

\end{abstract}


\maketitle

\section{Introduction.}

Effective field theories (EFTs) now form a standard tool in nuclear physics
\cite{Beane:2000fx,Bedaque:2002mn,Epelbaum:2008ga}. They rely on a separation 
between the momentum scales of interest in low-energy nuclear physics and those of the underlying physics of quantum chromodynamics. When this separation of scales is large enough, physical observables can be expanded as power series in ratios of low-energy to high-energy scales. The resulting theories then provide systematic frameworks for building these expansions.

A key element of any EFT is the power counting that quantifies the importance of the terms in the theory, both as they appear in the Lagrangian or Hamiltonian and as they contribute to observables. A general tool for determining the power counting is the renormalisation group (RG) \cite{Wilson:1973jj}. This introduces a cut-off or subtraction scale to regulate the theory. This scale should lie between the scales of interest and those of the underlying physics but its exact value is a matter of choice and hence observables should be independent of it. By analysing the cutoff dependence
of terms in the theory, we can determine their scaling behaviour. This can be mapped onto a power counting for their contributions to observables.

The first applications of these ideas to nuclear physics 
\cite{Kaplan:1998tg,vanKolck:1998bw,Gegelia:1998gn,Birse:1998dk} showed how an EFT based on contact interactions could reproduce the effective-range expansion 
\cite{Bethe:1949yr}. The extension to include pion-exchange forces remains a subject of debate \cite{Epelbaum:2009sd,Birse:2010fj} but the ``pionless EFT" is a well-defined theory that can be applied to few-nucleon systems with momenta well below the pion mass. It is based on an expansion in powers of ratios of momenta to the pion mass. The leading term in this expansion is a fixed point of the RG that describes the unitary limit of two-body scattering \cite{Braaten:2004rm}. (Other, more unstable, fixed points also exist, but it is not clear whether these have any physical relevance \cite{Birse:2015iea}.)

The pionless EFT was extended to describe electromagnetic (EM) couplings in Refs.~\cite{Kaplan:1998sz,Chen:1999tn}. In the case of the deuteron charge form factor,
the first three terms in the small-scale expansion follow from the effective-range expansion, as has long been known \cite{Sprung:1990zz,Bhaduri:1990cj,Kermode:1991zz}. The first new low-energy constant (LEC) appears at third order. This describes short-range physics that is not encoded in the potential and so it must be fixed using an EM observable.
The orders at which new terms like this appear in observables follow from the $1/r$ enhancement of the short-distance wave function in the unitary limit \cite{Birse:2010fj,Valderrama:2014vra}.

In this work, we apply a Wilsonian RG to the EM couplings of two particles interacting via $S$-wave contact interactions. For simplicity, we focus on systems 
with a total spin of zero.
This extends the work of Ref.~\cite{Birse:1998dk} which analysed the RG for the potential, identifying the fixed point that governs the effective-range expansion and the scaling of perturbations around it. The approach relies on the  methods developed in Refs.~\cite{Kvinikhidze:1997wn,Kvinikhidze:1997wp,Kvinikhidze:1997gd} for constructing gauge-invariant currents by ``gauging equations". Originally developed for use in three-body systems, this method was extended to equations with cutoffs in Ref.~\cite{Kvinikhidze:2009be}. 

The starting point is a one-particle-irreducible 5-point amplitude \cite{Kvinikhidze:2007eu}, where the two particles are coupled to a single EM field or photon. This can be thought of as a bremsstrahlung amplitude, with couplings of the EM field to the external legs omitted. Demanding that the fully off-shell 5-point amplitude be independent of the cutoff leads to a flow equation for the two-body interaction current density. When rescaled to express all variables in units of the cutoff, this becomes the RG equation. For any fixed-point potential, this equation 
has a fixed-point solution for the corresponding interaction current. The scaling with the cutoff of perturbations around it can then be analysed to determine the power counting. Previous studies of RG equations for the interaction current 
\cite{Kvinikhidze:2007eu,Nakamura:2006hc} used a simpler prescription for gauging the cutoff, which does not satisfy the relevant Ward-Takahashi identity (WTI) 
\cite{Kvinikhidze:2009be}, and did not examine their scaling properties.

For the fixed-point potential describing the unitary limit, we find the corresponding fixed point of the interaction charge density. The perturbations around this are of two types. The first consists of ones that can be generated by gauging terms already present in the potential \cite{Siegert:1937yt}. In nuclear physics, such terms are often referred as arising from Siegert's theorem (see, for example, 
Ref.~\cite{Friar:1984zza}). In the case of the pionless EFT, they have coefficients that are determined by the effective-range expansion for the scattering amplitude 
\cite{Sprung:1990zz,Kermode:1991zz}. 
The second type consists of perturbations that satisfy a homogeneous version of the RG equation. These describe short-range physics that is not included in the potential. Their coefficients must therefore be fixed using EM observables. 

Our results for on-shell (energy-dependent) perturbations agree with those found previously using other methods \cite{Kaplan:1998sz,Chen:1999tn,Valderrama:2014vra}. The use of the Wilsonian RG means that they are obtained here in a general framework that can be extended to other cases. For example, we also examine off-shell (momentum-dependent) perturbations, and we show that these do not make independent contributions to the on-shell bremsstrahlung amplitude. This reinforces the principle that the off-shell form of any potential is not observable. Also, the use of a momentum cutoff rather than dimensional regularisation means that our approach is closer to standard methods used in nuclear physics.

This paper is structured as follows. In Section II, we derive a flow equation for the two-body interaction current density by demanding cutoff independence of the  off-shell 5-point amplitude. We then rescale this to put it in the form of an RG equation.
In Section III, we find the fixed-point solution to this equation that corresponds to the unitary limit. We then find the perturbations around this that satisfy the linearised version of the RG equation. The corresponding physical 5-point amplitude is constructed in Section IV, along with the form factor for the two-body bound state. Off-shell perturbations are considered in the Appendix.

\section{RG equation for the two-body interaction current}

\subsection{Cutoff independence of amplitudes}

We consider here scattering of two equal-mass particles by an interaction that acts only in $S$ waves and so does not depend on the directions of the momenta. We assume that the particles are either spinless or are in a state with total spin zero, so the system has no magnetic couplings.
An RG equation for the potential is obtained by imposing a cutoff on the Lippmann-Schwinger (LS) equation for the off-shell $K$ matrix,
\be
K(k',k;p)=V_\Lambda(k',k;p) +{\cal P}
\int \mathrm{d}^3{\bm u}\,V_\Lambda(k',u;p) \frac{M\theta(\Lambda-u)}{(2\pi)^3(p^2-u^2)} K(u,k;p),
\label{LS-K}
\ee
and demanding that  $K$ be independent of the cutoff $\Lambda$ \cite{Birse:1998dk}. Here $V_\Lambda(k',k;p)$ is the cut-off potential expressed as a function of off-shell momenta, $k$ and $k'$, and the on-shell momentum $p$.

This condition leads to a differential equation for the potential which can be written in the schematic form
\be
\frac{\partial V_\Lambda}{\partial \Lambda} +V_\Lambda G_0 \frac{\partial \theta}{\partial \Lambda} V_\Lambda=0,  \label{mike-RG}
\ee
where the standing-wave (principal-value) propagator is
\be
G_0(p,k)=\frac{M}{(2\pi)^3(p^2-k^2)}.  \label{G0}
\ee
The corresponding propagators with $\pm\mathrm{i}\epsilon$ prescriptions will be denoted below by 
 $G_0^\pm$. With the momentum variables written out explicitly, Eq.~(\ref{mike-RG}) reads 
\be
\frac{\partial V_\Lambda(k',k;p)}{\partial \Lambda} +V_\Lambda(k',\Lambda;p) 
\frac{M}{2\pi^2(p^2-\Lambda^2)} V_\Lambda(\Lambda,k;p)=0,
\label{RG-in-var}
\ee
after integrating over angles.

An analogous equation for the two-body interaction current can be obtained by applying a similar procedure to the one-particle-irreducible 5-point amplitude \cite{Kvinikhidze:2007eu}. It can be obtained from the $K$ matrix by using the ``gauging of equations" method, as extended to equations with cutoffs in Ref.~\cite{Kvinikhidze:2009be}. Applied to the $K$ matrix above, it yields the off-shell 5-point amplitude
\be
K^\mu=(1+K G_0\theta) V_\Lambda^\mu(1+ G_0\theta K) +K(G_0\theta)^\mu K,   \label{Gtmu}
\ee
where $\mu$ is a Lorentz index and $V_\Lambda^\mu(\bm k',\bm k; \bm q; p',p)$ denotes the interaction current. Note that in this work we write Lorentz indices on amplitudes and interactions as superscripts. Other labels, including those for fixed points, orders in momentum and RG eigenvalues, are
all denoted by subscripts.
For simplicity, we work in the Breit frame, where the initial and final total momenta are $-\bm q/2$ and $+\bm q/2$ respectively, where $\bm q$ is the momentum transfer. In expressions like (\ref{Gtmu}) we do not make explicit the on-shell variables $E$ and $E'$, or equivalently $p$ and $p'$. Hence, for example, an expression like $BA^\mu C$ should be interpreted as
\be
BA^\mu C\rightarrow B(p')A^\mu(p',p)C(p).   \label{E'E-E}
\ee

Here $(G_0\theta)^\mu$ is the gauged propagator, including the cutoff used to regulate the loop integral in the LS equation (\ref{LS-K}). We define this using the symmetric prescription
\be
(G_0\theta)^\mu=\frac{1}{2}(\theta G_0^\mu+G_0^\mu\theta +\theta^\mu G_0+G_0\theta^\mu),
\label{GI-Lam}
\ee
where $\theta^\mu$ is the gauged step function introduced in Ref.~\cite{Kvinikhidze:2009be}. Here the gauged free propagators are 
\be\label{GI-props}
G_0^{+\mu}=G_0^+\Gamma^\mu G_0^+,\hspace{1cm}G_0^\mu=
\frac{1}{2}\left(G_0^+\Gamma^\mu G_0^++G_0^-\Gamma^\mu G_0^-\right),
\ee
where $\Gamma^\mu$ denotes the one-body current, 
\be
\Gamma^\mu=-(G_0^{-1})^\mu.
\ee

We note that it is important to apply the gauging prescription to the cutoff as well as the propagator in this expression \cite{Kvinikhidze:2009be}, as otherwise the interaction current fails to satisfy the WTI,
\begin{align}
q_\mu V_\Lambda^\mu(\bm k',\bm k; \bm q;p',p) =&z_1 V_\Lambda(|\bm k'-\bm q/2|,k;p)  
+z_2  V_\Lambda(|\bm k'+\bm q/2|,k;p) \nn
& -  z_1V_\Lambda(k',|\bm k+\bm q/2|;p') 
-  z_2 V_\Lambda(k',|\bm k-\bm q/2|;p'),  \label{WTI-V}
\end{align}
where $z_{1,2}$ are the charges of the two particles. The prescription used in earlier works \cite{Kvinikhidze:2007eu,Nakamura:2006hc}, which corresponds to $(G_0\theta)^\mu=\theta G_0^\mu\theta $, does not satisfy this requirement. 

Requiring $\Lambda$ independence of the 5-point amplitude, Eq.~(\ref{Gtmu}) and using the LS equation (\ref{LS-K}), we obtain the evolution equation for the interaction current $V^\Lambda_\mu$:
\be
\frac{\partial V_\Lambda^\mu}{\partial\Lambda}
+V_\Lambda G_0\frac{\partial\theta}{\partial\Lambda}V_\Lambda^\mu+V_\Lambda^\mu G_0\frac{\partial\theta}{\partial\Lambda} V_\Lambda
+V_\Lambda\frac{\partial(G_0\theta)^\mu}{\partial\Lambda} V_\Lambda =0.
 \label{5rg}
\ee
Note that this equation can also be obtained formally by applying the gauging prescription directly
to the evolution equation for the potential (\ref{mike-RG}) and using $(G_0\partial\theta/\partial\Lambda)^\mu=\partial(G_0\theta)^\mu/\partial\Lambda$.

Since we consider only spinless systems here, gauge invariance allows us to construct the space components of the current ${\bm V}_\Lambda$ from the interaction charge density $V^0_\Lambda$ by using the WTI, Eq.~(\ref{WTI-V}). A further simplification follows for energy-independent regulators since the time component of $(G_0\theta)^\mu$ does not contain a piece from gauging the step function, as $\theta^0=0$. In this case and working in the Breit frame, the gauged propagator takes the form
\begin{align}
\la\bm k'| (G_0\theta)^0|\bm k\ra
&=\la\bm k'| \frac{1}{2}(\theta G_0^0+G_0^0\theta )|\bm k\ra \nn
&= 
\frac{M^2}{2(2\pi)^3}\,\frac{ \theta(\Lambda-k')+ \theta(\Lambda-k)}{(p'^2-k'^2)(p^2-k^2)}\bigl[z_1\delta(\bm k'-\bm k+\bm q/2)+z_2\delta(\bm k'-\bm k-\bm q/2)\bigr].
\label{GI-Lam-0}
\end{align}
Since our interaction acts only in $S$ wave, we need the matrix element of this quantity in states with relative $l=0$,
\be
\la k'|\frac{\partial(G_0\theta)^0}{\partial\Lambda}|k\ra =\frac{ZM^2}{4(2\pi)^4qk' k} \,\frac{[\delta(\Lambda-k')+\delta(\Lambda-k)]}{(p'^2-k'^2)(p^2-k^2)}  \,
\theta\!\left[\left(k+\frac{q}{2}\right)^2-k'^2\right]
\theta\!\left[k'^2-\left(k-\frac{q}{2}\right)^2\right],  \label{s-wave-me}
\ee
where $Z=z_1+z_2$ is the total charge of the system.

\subsection{Rescaling and RG equations}\label{resc-vecV}

The evolution equations in the previous section can be put into the standard form of RG equations by rescaling to express them in terms of dimensionless variables, as described in Ref.~\cite{Birse:1998dk}. This makes it possible to identify the fixed-point solutions that describe scale-free physical systems. The dependence on the cutoff $\Lambda$ of perturbations around these solutions can then be used to determine the power counting that specifies the importance of their contributions to observables, such as the scattering or bremsstrahlung amplitudes, or the EM form factor.

We express all momentum variables in units of $\Lambda$,
\be
\hat k=k/\Lambda,\;  \hat k'=k'/\Lambda,\; \hat p=p/\Lambda,\; \mbox{etc.},
\ee
and we define the rescaled potential
\be
\widehat V_\Lambda=\frac{M\Lambda}{2\pi^2}\,V_\Lambda.
\ee
This satisfies the RG equation \cite{Birse:1998dk}
\be
\left(\Lambda\frac{\hat\partial}{\partial\Lambda}-\hat{k}'\frac{\partial}{\partial\hat{k}'}-\hat{k}\frac{\partial}{\partial\hat{k}}-\hat{p}'\frac{\partial}{\partial\hat{p}'}-1\right) 
\widehat V_\Lambda(\hat k',\hat k;\hat p)
+\widehat V_\Lambda(\hat k',1;\hat p)\,\frac{1}{\hat p^2-1}\,\widehat V_\Lambda(1,\hat k;\hat p)=0,
\label{RGE-pot}
\ee
where the partial derivative with respect to $\Lambda$ should now be understood to be taken for fixed \emph{rescaled} momentum variables.

To obtain a similar equation for the interaction charge density, we define the rescaled density
\be
\widehat V^0_\Lambda=\frac{\Lambda^3}{2\pi^2}\,V^0_\Lambda.
\label{rescV0}
\ee
The corresponding expression for the rescaled space components of the interaction current is 
\be
\widehat {\bm V}_\Lambda=\frac{M\Lambda^2}{2\pi^2}\,{\bm V}_\Lambda.
\label{rescVi}
\ee
The different rescalings needed to make these dimensionless can be understood from the gauging procedure. To obtain a density from an amplitude involves dividing it by a factor of the energy ($p^2/M$) whereas a current density involves dividing by a factor of momentum.

Using the expression (\ref{s-wave-me}) in the evolution equation (\ref{5rg}) and rescaling all momentum variables, we arrive at the RG equation for $\widehat V^0_\Lambda$,
\begin{align}
&\left(\Lambda\frac{\hat\partial}{\partial\Lambda}-\hat{k}'\frac{\partial}{\partial\hat{k}'}-\hat{k}\frac{\partial}{\partial\hat{k}} -\hat{q}\frac{\partial}{\partial\hat{q}}
-\hat{p}'\frac{\partial}{\partial\hat{p}'}-\hat{p}\frac{\partial}{\partial\hat{p}}-3\right)
 \widehat V_\Lambda^0(\hat{k}',\hat{k};\hat q;\hat p',\hat p)
\nn
&\quad+\widehat V_\Lambda(\hat{k}',1;\hat p') \frac{1}{\hat p'^2-1}
\widehat{V}_\Lambda^0(1,\hat{k};\hat q;\hat p',\hat p)
+\widehat{V}_\Lambda^0(\hat{k}',1;\hat q;\hat p',\hat p) \frac{1}{\hat p^2-1} 
\widehat V_\Lambda(1,\hat{k};\hat p) \nn
&\quad+\frac{Z}{2 \hat q}\, \frac{\widehat V_\Lambda(\hat k',1;\hat p') }{\hat p'^2-1}
\int^{1+\frac{\hat q}{2}}_{|1-\frac{\hat q}{2}|} \hat u\,\mathrm{d}\hat u     
\frac{\widehat V_\Lambda(\hat u,\hat k;\hat p)}{\hat p^2-\hat u^2}\nn
&\quad+\frac{Z}{2 \hat q} \int^{1+\frac{\hat q}{2}}_{|1-\frac{\hat q}{2}|} \hat u\,\mathrm{d}\hat u
\frac{\widehat V_\Lambda(\hat k',\hat u;\hat p') }{\hat p'^2-\hat u^2}  \,   
\frac{\widehat V_\Lambda(1,\hat k;\hat p)}{\hat p^2-1}=0.\label{eqV0-expl}
\end{align}

\section{Fixed points and their perturbations}

\subsection{Fixed points}

Fixed points are solutions to these equations that are independent of $\Lambda$. In the case of the short-range potential, these have been well studied using various approaches \cite{Kaplan:1998tg,vanKolck:1998bw,Gegelia:1998gn,Birse:1998dk}. 
There is, of course, the trivial solution, $\widehat V=0$. The scaling behaviour of perturbations around this point is governed by naive dimensional analysis. The same holds for the terms in the interaction density and so we do not consider the expansion around this fixed point further here.

In addition, there are various nontrivial fixed points, as discussed in Ref.~\cite{Birse:2015iea}. All of these are unstable, having at least one relevant perturbation, and so they describe fine-tuned systems. The RG equation (\ref{eqV0-expl}) can be used to find fixed points of the charge density corresponding to each of these. However, in practice, only the one with a single unstable direction is of physical relevance. 
This is the fixed point that describes scattering in the unitary limit \cite{Braaten:2004rm}, and which provides the starting point for the effective-range expansion of the scattering amplitude \cite{Bethe:1949yr}. For the sharp momentum cutoff used here, this depends only on the energy, not the off-shell momenta, and it has the form \cite{Birse:1998dk}
\be
\widehat V_F(\hat p)=\left[
\int\frac{\theta(1- \hat k) }{\hat p^2-\hat k^2}\,\hat k^2 \mathrm{d}\hat k\right]^{-1}=
-\left[1-\frac{\hat p}{2}\ln\frac{1+\hat p}{1-\hat p}\right]^{-1},  \label{gen-hatV**}
\ee
This satisfies the boundary condition that it be analytic in the energy, or $\hat p^2$, for small $\hat p$.

By substituting $V_F$ into the RG equation (\ref{eqV0-expl}) for the interaction charge density, we get a differential equation for a fixed-point solution, $\widehat V^0_F(\hat q;\hat p',\hat p)$, which is also independent of the off-shell momenta:
\begin{align}\label{eq-fix-V0-expl}
&\left(
 -\hat{q}\frac{\partial}{\partial\hat{q}}
-\hat{p}'\frac{\partial}{\partial\hat{p}'}-\hat{p}\frac{\partial}{\partial\hat{p}}-3\right)
\widehat V_F^0(\hat q;\hat p',\hat p)
+\left( \frac{\widehat V_F(\hat p')}{\hat p'^2-1}
+ \frac{\widehat V_F(\hat p)}{\hat p^2-1}  \right) \widehat V_F^0(\hat q;\hat p',\hat p) \nn
&\quad +\frac{ Z}{ 2\hat q}\, \frac{\widehat V_F(\hat p') \widehat V_F(\hat p)}{\hat p'^2-1}\int^{1+\frac{\hat q}{2}}_{|1-\frac{\hat q}{2}|}     
\frac{\hat ud\hat u }{\hat p^2-\hat u^2}+(p'\leftrightarrow p) =0.
\end{align}
The solution to this equation that is analytic in $\hat p^2$ can be written in the form
\be\label{V0-fp}
\hat V^0_F(\hat q;\hat p',\hat p)
=-\widehat V_F(\hat p')\la (\widehat{G_0\theta})^0\ra \widehat V_F(\hat p),
\ee
where the integrated and rescaled gauged propagator is
\be
\la (\widehat{G_0\theta})^0\ra=\frac{2\pi^2\Lambda}{M^2}\la (G_0\theta)^0\ra
=\frac{Z\Lambda}{8\pi}\left[\int \mathrm{d}^3\bm u\,
\frac{\theta(\Lambda-u)}{(p'^2-[\bm u+\bm q /2]^2)(p^2-\bm u^2)}+(p'\leftrightarrow p)\right] .
\ee
Expressing the fixed-point charge density in this form shows that it could also be obtained directly by gauging the (energy-dependent) fixed-point potential.

The corresponding current density satisfies an RG equation similar to Eq.~(\ref{eqV0-expl}). However, as noted above, it can also be found using the WTI, Eq.~(\ref{WTI-V}). For the fixed point, this takes the form
\be
q_\mu V_F^\mu(\bm q;p',p)=Z[V_F(p)-V_F(p')].  \label{wti-F}
\ee
Using the fact that the energy transfer in the Breit frame  is $q^0=(p'^2-p^2)/M$,
it is easy to see that the current density
\be
\bm V_F(\bm q;p',p)=\left[\frac{p'^2-p^2}{M}\,V^0_F(q;p',p)-Z[V_F(p)-V_F(p')]\right] 
\frac{\bm q}{q^2}.
\label{bf-current}
\ee
satisfies this WTI. After rescaling according to Eq.~(\ref{rescVi}), this becomes
\be
\widehat{\bm V}_F(\hat{\bm q};\hat p',\hat p)=\left[(\hat p'^2-\hat p^2)
\widehat V_F^0(\hat q;\hat p',\hat p)
- Z(\hat V_F(\hat p)-\hat V_F(\hat p')) \right]\frac{\hat{\bm q}}{\hat q^2}.\label{Vi-F}
\ee
Although the factor of $\hat q^{-2}$ looks nonanalytic in $\hat q$, this is cancelled by factors of $\hat q$ in the expression in brackets:
\be
\hat V_F^0(\hat q;\hat p',\hat p)-Z\,\frac{\hat V_F(\hat p)-\hat V_F(\hat p')}{\hat p'^2-\hat p^2}
= Z\hat V_F(\hat p') \frac{ \hat q^2}{36} \hat V_F(\hat p) +{\cal O}(\hat q^2\hat p^2)  \label{sim-q2*}
\ee
Finally, we should point out that the expression (\ref{bf-current}) for the current density applies to the Breit frame. In a general frame, Galilean invariance requires an additional term of the form
\be
\delta \bm V_F(\bm P',\bm P;p',p)=\frac{\bm P+\bm P'}{4m}\,V_F^0(q;p',p),
\ee
where $\bm P$ and $\bm P'$ are the initial and final centre-of-mass momenta.

\subsection{Linearised perturbations and scaling}\label{scaling}

The fixed point constructed above for the interaction charge density, like the potential it was obtained from, describes systems at the unitary limit. To describe more general systems, we need to add perturbations around these fixed points. It is convenient to expand these in terms of the eigenfunctions of the linearised RG equation, since these scale with definite powers of $\Lambda$
\cite{Birse:1998dk}.

We add a small perturbation $\Phi_\Lambda$ to the fixed-point potential,
\be
\widehat V_\Lambda(\hat{k}',\hat{k};\hat p)=
\widehat V_F(\hat p)+\Phi_\Lambda(\hat{k}',\hat{k};\hat p),
\ee
and we expand it in the form
\be
\Phi_\Lambda(\hat k',\hat k;\hat p)=\sum_\nu C_\nu\Lambda^\nu\phi_\nu(\hat k',\hat k;\hat p).,
\ee
where the $\phi_\nu$ are eigenfunctions of the linearised RG equation obtained from
Eq.~(\ref{RGE-pot}),
\begin{align}\label{eqV-Ric-lin*}
&\left(\hat{k}'\frac{\partial}{\partial\hat{k}'}+\hat{k}\frac{\partial}{\partial\hat{k}}
+\hat{p}\frac{\partial}{\partial\hat{p}}+1\right)  \phi_\nu(\hat{k}',\hat{k};\hat p)
\nn
&\quad-\widehat V_F(\hat p') \frac{1}{\hat p'^2-1}\phi_\nu(1,\hat{k};\hat p)
-\phi_\nu(\hat{k}',1;\hat p) \frac{1}{\hat p^2-1}\widehat V_F(\hat p) 
=\nu\,\phi_\nu(\hat{k}',\hat{k};\hat p),
\end{align}
with eigenvalue $\nu$ \cite{Birse:1998dk}. Here we focus on the energy-dependent 
eigenfunctions,
\be
\phi _{2n-1}(\hat p)=\hat p^{2n}\widehat V_F^2(\hat p),  \label{en-pertn}
\ee
which have scaling eigenvalues $\nu=2n-1=-1,\ 1,\ 3,\ \dots$. Terms that depend on off-shell momenta are discussed in Appendix \ref{off-shell}.

In the same way, we perturb the charge density,
\be
\widehat V_\Lambda^0(\hat{k}',\hat{k};\hat q;\hat p',\hat p)=
\widehat V_F^0(\hat q;\hat p',\hat p)+\Phi_\Lambda^0(\hat{k}',\hat{k};\hat q;\hat p',\hat p),
\label{sm-pert-0-f}
\ee
and expand it as
\be
\Phi^0_\Lambda(\hat k',\hat k;\hat q;\hat p',\hat p)=\sum_\nu D_\nu \Lambda^\nu
\phi^0_\nu(\hat k',\hat k;\hat q;\hat p',\hat p).
\ee

The eigenfunctions here satisfy the linearised RG equation,
\begin{align}\label{eqV0-lin}
&\left(\hat{k}'\frac{\partial}{\partial\hat{k}'}+\hat{k}\frac{\partial}{\partial\hat{k}}
+\hat{q}\frac{\partial}{\partial\hat{q}}
+\hat{p}'\frac{\partial}{\partial\hat{p}'}+\hat{p}\frac{\partial}{\partial\hat{p}}+3
\right)  
\phi^0_\nu(\hat{k}',\hat{k};\hat q;\hat p',\hat p)\nn
&\quad -\widehat V_F(\hat p') \frac{1}{\hat p'^2-1}\phi_\nu^0(1,\hat{k};\hat q;\hat p',\hat p)
-\phi_\nu^0(\hat{k}',1;\hat q;\hat p',\hat p) \frac{1}{\hat p^2-1}\widehat V_F(\hat p) \nn
&\quad -\phi_\nu(\hat{k}',1;\hat p') \frac{1}{\hat p'^2-1}\widehat V_F^0(\hat q;\hat p',\hat p)
-\widehat V_F^0(\hat q;\hat p',\hat p) \frac{1}{\hat p^2-1} \phi_\nu(1,\hat{k};\hat p) 
\nn
&\quad -\frac{Z}{ 2\hat q}\, \frac{\widehat V_F(\hat p') }{\hat p'^2-1}
\int^{1+\frac{\hat q}{2}}_{|1-\frac{\hat q}{2}|}      
\frac{\phi_\nu(\hat u,\hat k;\hat p)}{\hat p^2-\hat u^2}\,\hat u\,\mathrm{d}\hat u
-\frac{Z}{ 2\hat q}\, \frac{\phi_\nu(\hat k',1;\hat p') }{\hat p'^2-1}
\int^{1+\frac{\hat q}{2}}_{|1-\frac{\hat q}{2}|} \frac{\widehat V_F(\hat p)}{\hat p^2-\hat u^2}
\,\hat u\,\mathrm{d}\hat u\nn
&\quad -\frac{ Z}{2 \hat q} \int^{1+\frac{\hat q}{2}}_{|1-\frac{\hat q}{2}|} 
\frac{\widehat V_F(\hat p') }{\hat p'^2-\hat u^2}\,\hat u\,\mathrm{d}\hat u\,
\frac{\phi_\nu(1,\hat k;\hat p)}{\hat p^2-1}
-\frac{ Z}{2 \hat q} \int^{1+\frac{\hat q}{2}}_{|1-\frac{\hat q}{2}|} 
\frac{\phi_\nu(\hat k',\hat u;\hat p') }{\hat p'^2-\hat u^2} \,\hat u\,\mathrm{d}\hat u\,
\frac{\widehat V_F(\hat p)}{\hat p^2-1} \nn
&\qquad=\nu \phi^0_\nu(\hat{k}',\hat{k};\hat q;\hat p',\hat p).
\end{align}
This equation is coupled to Eq.~(\ref{eqV-Ric-lin*}) by the terms in lines 3 to 5. It therefore has two kinds of solution. One consists of solutions driven by $\Phi_\Lambda$ through the terms that couple Eq.~(\ref{eqV0-lin}) to (\ref{eqV-Ric-lin*}). Since these are driven by the $\phi_\nu$, their coefficients should match: $D_\nu=C_\nu$. The other comprises solutions to the homogeneous version of Eq.~(\ref{eqV0-lin}), where $\phi_\nu$ is set to zero. These have independent coefficients $D_\nu$.

The solution driven by $\phi_{2n-1}$ is momentum-independent and has the form
\be
\phi^0_{2n-1}(\hat q;\hat p',\hat p)=\widehat V_F^0(\hat q;\hat p',\hat p)
\left( \hat p^{2n}\widehat V_F(\hat p)+ \hat p'^{2n}\widehat V_F(\hat p')\right)
-Z\widehat V_F(\hat p')\, \frac{\hat p'^{2n}-\hat p^{2n}}{\hat p'^2- \hat p^2}\,
\widehat V_F(\hat p),
\label{phi0nu}
\ee
where the final term is absent for $n=0$. The corresponding eigenvalue is,
of course, the same as for $\phi_{2n-1}$: $\nu=2n-1$. Like $\widehat V^0_F(\hat q;\hat p',\hat p)$, this contribution to the charge density could also be obtained directly by gauging the corresponding term in the potential, in this case $\phi_{2n-1}(\hat p)$. Note that, when gauging Eq.~(\ref{en-pertn}), we need to write it in the form $\widehat V_F(\hat p)\hat p^{2n}\widehat V_F(\hat p)$ and to be careful to maintain the ordering of the operators in this expression. The resulting three terms in the density are the gauged versions of the three factors in Eq.~(\ref{en-pertn}).

Then we have the solutions to the homogeneous equation. The ones that do not depend on off-shell momenta have the form
\be
\phi^0_{lmn}(\hat q;\hat p',\hat p)
=\widehat V_F(\hat p')\hat q^{2l+2}\,\hat p'^{2m}\,\hat p^{2n}\widehat V_F(\hat p),
\label{phi0lmn}
\ee
with eigenvalues $\nu=2(l+m+n)+3$ for $n,l,m=0,1,2,\dots$. The extra power of $\hat q^2$ is needed to make the space components of the current regular as $\hat q\rightarrow 0$, as discussed below.
For a Hermitian current operator, the coefficients must satisfy $D_{lmn}=D_{lnm}$. These coefficients are the LECs that describe short-range physics that is not encoded in the potential and that must be fixed using EM observables. Again we leave discussion of off-shell terms to Appendix \ref{off-shell}.

As described above, the space part of the current density can be constructed using the WTI, Eq.~(\ref{WTI-V}). For the terms that arise from gauging energy-dependent perturbations in the potential, the currents ${\bm\phi}_{2n-1}$ have a form similar to the fixed-point current, Eq.~(\ref{Vi-F}).
For the potential-independent perturbations to the charge density in Eq.~(\ref{phi0lmn}), the WTI leads to the Breit-frame current density,
\be
\bm\phi_{lmn}(\hat q;\hat p',\hat p)= (\hat p'^2-\hat p^2)\,\phi^0_{lmn}(\hat q;\hat p',\hat p)\,
\frac{\hat{\bm q}}{\hat q^2} .
\ee
We see that the additional factor of $\hat q^2$ in the charge density $\phi^0_{lmn}$ cancels the factor of $\hat q^{-2}$ in this expression and gives a current density that is analytic in $\hat q$ for small $\hat q$.

\section{Observables}

\subsection{Scattering amplitudes}

The scattering amplitude obtained from the fixed-point potential 
Eq.~(\ref{gen-hatV**}) has the form $T(p)=4\pi/\mathrm{i}Mp$.
This is of order $Q^{-1}$, where $Q$ denotes a generic low-energy scale. 
As pointed out in Ref.~\cite{Birse:1998dk}, the power counting is directly 
related to the scaling behaviour of terms in the rescaled potential, 
with a term scaling as $\Lambda^\nu$ contributing at order $Q^{\nu-1}$. 

The interaction current density for this could be used to find the corresponding 
5-point amplitude. However, this fixed point describes a system where a bound 
state has just merged with the continuum and so its form factor is ill-defined.
Also, this fixed point is unstable, describing a highly fine-tuned system. It is therefore more useful to start from a potential where the leading, energy-independent perturbation is included to all orders. This describes systems with a large but finite scattering length $a$ and, for $a>0$, a low-energy bound state. 

With $C_{-1}$ included to all orders, the potential has the form
\be
\widehat V_{S\Lambda}(p)=\left[\widehat V_F(p)^{-1}-C_{-1}\Lambda^{-1}\right]^{-1}. \label{pot-fsl}
\ee
It can be shown to satisfy the RG equation (\ref{RGE-pot}) most easily by rewriting that equation in the form of a linear equation for $V^{-1}$ \cite{Barford:2002je}.
The resulting $T$ matrix is, in unscaled form,
\be\label{t-Ma}
T_S(p)=\frac{4\pi}{M}\,\frac{1}{\mathrm{i}p-2C_{-1}/\pi}
=\frac{4\pi}{M}\,\frac{1}{\mathrm{i}p+1/a},
\ee
showing that $C_{-1}$ is related to the scattering length by
$C_{-1}=-\pi/2 a$. If $a>0$, this has a bound-state pole at $E=-E_B=-1/Ma^2$.

One way to interpret this potential is to say that $1/a$ has been added to the list of low-energy scales, which is needed for momenta $p\gtrsim |1/a|$. The term $C_{-1}$ could then be rescaled to make it part of the fixed point. The resulting $T$ matrix is of order $Q^{-1}$ where $Q$ in this case is either $p$ or $1/a$.

The interaction charge density for the potential Eq.~(\ref{pot-fsl}) is
\be
\widehat V^0_{S\Lambda}(\hat q;\hat p',\hat p)
=-\widehat V_{S\Lambda}(\hat p')\la (\widehat{G_0\theta})^0\ra 
\widehat V_{S\Lambda}(\hat p).
\ee
Like Eq.~(\ref{V0-fp}), this can be obtained either by solving the RG equation or by simply gauging the potential $\widehat V_{\Lambda}(p)$. The irreducible 5-point amplitude is given by the scattering-wave analogue of Eq.~(\ref{Gtmu}),
\be
T^0=(1+T G^+_0\theta) V_\Lambda^0(1+ G^+_0\theta T) 
+T\la (G^+_0\theta)^0 \ra T.
\label{T0}
\ee
Inserting $V^0_{S\Lambda}(p)$ into this expression and using the LS equation for $T$, we get
\be
T_S^0(q;p',p)=T_S(p')\left[\la (G^+_0\theta)^0\ra
-\la ( G_0\theta)^0\ra\right] T_S(p).
\label{ga-t-Ma}
\ee
Here we have taken advantage of the fact that, for a momentum-independent potential, these are algebraic rather than integral equations.

The angle-integrated gauged propagators can be evaluated with the help of 
Eq.~(\ref{GI-props}), giving
\begin{align}
\la (G^+_0\theta)^0\ra -\la (G_0\theta)^0\ra 
&=\frac{1}{2}\left(\la G_0^+\Gamma^0 G_0^+\ra - \la G_0^-\Gamma^0 G_0^-\ra\right)\nn
&=-\mathrm{i}\,\frac{ZM^2}{4\pi q} \ln\frac{p+p'-q/2}{p+p'+q/2}\,.   \label{p'+p-q2}
\end{align}
Note that if the argument of the logarithm is not real and positive, the imaginary part should be defined such that it vanishes as $q\rightarrow 0$.
The resulting expression for the 5-point amplitude is
\be\label{t0-LO}
T_S^0(q;p',p)=-T_S(p')\left[\mathrm{i}\,\frac{ZM^2}{4\pi q} 
\ln\frac{p+p'-q/2}{p+p'+q/2}\right] T_S(p).
\ee
In terms of low-energy scales, this expression is of order $Q^{-3}$. This is as expected since the scattering amplitude is of order $Q^{-1}$ and gauging any quantity removes one power of energy ($p^2$), replacing it by a coupling to the EM potential.

At higher orders, the charge density includes the perturbations in 
Eqs.~(\ref{phi0nu},\ref{phi0lmn}). After resumming the contributions of $C_{-1}$ so that $V_F$ is replaced by $V_S$, the current density has the form
\begin{align}
V^0_\Lambda(q;,p',p)=&\, V^0_{S\Lambda}(q;,p',p)
+\frac{ZM}{2\pi^2}\sum_{n=1}^\infty C_{2n-1}\biggl[V^0_{S\Lambda}(q;,p',p)
\Bigl(p^{2n}V_{S\Lambda}(p)+ p'^{2n}V_{S\Lambda}(p')\Bigr)\nn
&\qquad\qquad\qquad\qquad\qquad\qquad\qquad -\left.M\,V_{S\Lambda}(p')\, 
\frac{p'^{2n}-p^{2n}}{p'^2- p^2}\,V_{S\Lambda}(p)\right]\nn
&+\frac{M^2}{2\pi^2}\,V_{S\Lambda}(p')
\sum_{l,m,n=0}^\infty D_{lmn}\,q^{2l+2}\,p'^{2m}\,p^{2n}\,V_{S\Lambda}(p).
\end{align}
The 5-point amplitude to first order in the $C_\nu$ can be obtained by inserting this into Eq.~(\ref{T0}). The pieces arising from the first two terms containing factors of $V^0_{S\Lambda}$ can be expressed in the form 
$-\delta T\la ( G_0\theta)^0\ra T_S-(p'\leftrightarrow p)$, where $\delta T(p)$ denotes the contribution to the scattering amplitude from the 
$C_\nu$ terms in the potential,
\be
\delta T(p)=\frac{M}{2\pi^2}\,T_S(p)\sum_{n=1}^\infty C_{2n-1}p^{2n}\,T_S(p).
\ee
Combining this with the first-order pieces of the final term in Eq.~(\ref{T0}), we get a contribution of the form $\delta T[\la (G^+_0\theta)^0\ra-\la ( G_0\theta)^0\ra] T_S+(p'\leftrightarrow p)$, which can be evaluated using Eq.~(\ref{p'+p-q2}). The full 5-point amplitude at this order is thus:
\begin{align}
T^0(q;p',p)=&\, T_S^0(q;p',p)-\delta T(p')\,\mathrm{i}\,\frac{ZM^2}{4\pi q} 
\ln\frac{p+p'-q/2}{p+p'+q/2}\, T_S(p)-(p'\leftrightarrow p)\nn
&-\frac{M^2}{2\pi^2}\, T_S(p')\sum_{n=1}^\infty C_{2n-1}\,
\frac{ p'^{2n}- p^{2n}}{( p'^2-  p^2)}\, T_S(p)\nn
&+\frac{M^2}{2\pi^2}\,T_S(p')\sum_{l,m,n=1}^\infty D_{lmn}\, 
q^{2l+2}\, p'^{2m}\, p^{2n}\, T_S(p).
\label{t0-linear}
\end{align}

Remembering that $T_S(p)$ is of order $Q^{-1}$, we see that the $p^{2n}$ term in the potential contributes to the 5-point amplitude at order $Q^{2n-4}$. Similarly the 
potential-independent term in the charge density proportional to $q^{2l+2}\, p'^{2m}\, p^{2n}$ appears at order $Q^{2(n+l+m)}$. All of these are shifted two orders lower in the power counting compared to naive dimensional analysis. This behaviour can be understood more directly as a consequence of the $1/r$ form of the short-distance wave functions in the unitary limit \cite{Birse:2010fj,Valderrama:2014vra}, which enhances the contributions of all short-range operators. The first contribution that is not just a result of gauging the potential ($D_{000}\,q^2$) appears at order $Q^0$, as noted in Refs.~\cite{Kaplan:1998sz,Chen:1999tn}. (Since the leading amplitude is of order $Q^{-3}$, this term is of order N$^3$LO in the notation of those papers.)

The power counting for these contributions can be obtained from the scaling with $\Lambda$ of the corresponding terms in the rescaled charge density. A term in the operator that scales as $\Lambda^\nu$ contributes to the amplitude at order $Q^{\nu-3}$. The difference of three orders between the scaling with $\Lambda$ and the power counting in low-energy scales $Q$ follows from the rescaling of the charge density in Eq.~(\ref{rescV0}). The space components of the current density for an operator scaling as $\Lambda^\nu$ contribute at order $Q^{\nu-2}$, reflecting the different rescaling required, Eq.~(\ref{rescVi}).

\subsection{Effective-range expansion for EM amplitude}

Sor far, we have considered perturbative expansions of operators and amplitudes around the unitary limit. However, particularly for calculations of bound-state properties, this is not a convenient framework. Beyond leading order, the perturbative expansion of $T(p)$ in powers of the $C_\nu$ contains double and higher poles at the leading-order bound-state energy. These just reflect the shift in that energy that results from these perturbations. It is therefore more convenient to work with the effective-range expansion of $T(p)^{-1}$,
\be
T(p)^{-1}=\frac{M}{4\pi}\left[\mathrm{i}p+\frac{1}{a}
-\frac{2}{\pi}\sum_{n=1}^\infty C_{2n-1}\,p^{2n}\right],
\label{ERE-T}
\ee
where we have used the fact that the terms in this expansion are directly related to the energy-dependent purturbations around the unitary fixed point \cite{Birse:1998dk}.
For example, $C_1$ is given in terms of the effective range by $C_1=(\pi/4)r_e$. The low-energy bound state is given by the lowest-momentum zero of this expression, which we denote by $p=\mathrm{i}\gamma$.

In the case of the 5-point amplitude, we note that the leading order expression, Eq.~(\ref{t0-LO}), and the second line of Eq.~(\ref{t0-linear}) can both be obtained by applying the gauging-of-equations method to the corresponding terms in the effective-range expansion. Combining that method with the additional terms from Eq.~(\ref{phi0lmn}) gives a complete expression for the 5-point amplitude based on the effective-range expansion:
\begin{align}
T^0(q;p',p)=&-T(p')\left[\mathrm{i}\,\frac{ZM^2}{4\pi q} 
\ln\frac{p+p'-q/2}{p+p'+q/2}
+\,\frac{ZM^2}{2\pi^2}  \sum_{n=1}^\infty C_{2n-1}\,
\frac{ p'^{2n}- p^{2n}}{( p'^2-  p^2)}\right.\nn
&\qquad\qquad\quad -\left.\,\frac{M^2}{2\pi^2} \sum_{m,n,l=0} D_{lmn}\, 
q^{2l+2}\, p'^{2m}\, p^{2n} \right]T(p),
\label{ERE-T0}
\end{align}
where $T(p)$ is the full $T$ matrix from Eq.~(\ref{ERE-T}). The space components of this, $\bm T(q;p',p)$, can be determined from this with the aid of a WTI similar to Eq.~(\ref{WTI-V}).

Finally, the full bremsstrahlung amplitude consists of the irreducible amplitude plus terms where an EM field is coupled to one of the external legs:
\be
A^\mu=T^\mu-\Gamma^\mu\,G_0\,T-T\,G_0\,\Gamma^\mu.
\label{A-brems}
\ee

\subsection{Form factors}

For $a>0$, the $T$ matrix has a pole at the bound state energy. The charge form factor for this state can be determined from $T^0(q;p',p)$ by extracting its residue at the double pole, $E'=E=-E_B$. 

We start by defining a vertex function $\chi$ for the bound state by writing the $T$ matrix in the form 
\be\label{t-pole}
T(p)=\frac{\chi\chi^*}{E+E_B}+\dots,
\ee
where the dots denote pieces that are regular at $E=-E_B$. For the momentum-independent contact interactions we consider here, $\chi$ is simply a constant. The bound-state form factor $F(q)$ can then be defined from
\be
T^0(q;p',p)=\frac{\chi}{E'+E_B}\,F(q)\,\frac{\chi^*}{E-E_B}+\cdots,
\ee
where the dots denote pieces that are regular at either $E=-E_B$ or $E'=-E_B$.
This can be evaluated by rewriting it in the form
\be
F(q)=\left.\chi^* T(p')^{-1}T^0(q;p',p)T(p)^{-1}\chi\right|_{p=p'=\mathrm{i}\gamma}.
\label{ff-t0}
\ee

At leading order, we can find the normalisation simply by comparing Eqs.~(\ref{t-Ma}) and (\ref{t-pole}). This gives
\be
\chi^*\chi=\frac{8\pi}{M^2a},
\ee
since, at this order, $\gamma=\mathrm{i}/a$.
Using the corresponding expression, Eq.~(\ref{t0-LO}), for the 5-point amplitude, we get the form factor
\begin{align}
F(q)&=-\mathrm{i}\,\frac{2Z}{qa} 
\left.\ln\frac{p+p'-q/2}{p+p'+q/2}\right|_{p=p'=\mathrm{i}/a}\nn
&=\frac{4Z}{qa}\arctan\frac{qa}{4}.
\end{align}
Expanded in powers of the momentum transfer $q$, this becomes
\be
F(q)=Z\left(1-\frac{a^2}{48}\,q^2+\cdots\right),
\ee
with a charge radius given in terms of the scattering length as in 
Refs.~\cite{Sprung:1990zz,Bhaduri:1990cj,Kermode:1991zz}. The only scale in this form factor is the scattering length and so, taking $1/a$ to be of order $Q$, all terms are of order $Q^0$.

Going beyond leading order, the normalisation of the vertex function can be extracted from the effective-range expansion using
\be
(\chi^*\chi)^{-1}=M\lim_{p\rightarrow \mathrm{i}\gamma} 
\frac{T(p)^{-1}}{p^2+\gamma^2}
=\left.\frac{M}{2p}\,\frac{\mathrm{d}}{\mathrm{d}p} T(p)^{-1}
\right|_{p=\mathrm{i}\gamma}.
\ee
This gives
\be
(\chi^*\chi)^{-1}=\frac{M^2}{8\pi\gamma}\left[1-\frac{4}{\pi}\sum_{n=1}^\infty
(-1)^{n-1}n\,C_{2n-1}\,\gamma^{2n-1}\right].
\ee

Inserting the complete 5-point amplitude, Eq.~(\ref{ERE-T0}), in Eq.~(\ref{ff-t0}) gives the form factor from the extended effective-range expansion:
\begin{align}
F(q)=&\,\chi^*\chi\,\frac{M^2}{8\pi\gamma}\left[\frac{4Z\gamma}{q}
\arctan\frac{q}{4\gamma}
-\frac{4Z}{\pi} \sum_{n=1}^\infty (-1)^{n-1}n\,C_{2n-1}\,\gamma^{2n-1}\right.\nn
&\qquad\qquad +\left.\frac{4}{\pi} \sum_{l,m,n=0}^\infty (-1)^{m+n}D_{lmn}\,
q^{2l+2}\,\gamma^{2(m+n)+1}\right].
\end{align}
This expression extends to all orders the result from Ref.~\cite{Kaplan:1998sz}, where the pionless EFT was first applied to constructing the charge form factor from the effective-range expansion.
Expanding the arctangent in powers of $q$, we see that the $q$-independent terms in the numerator cancel with $\chi^*\chi$ to give a correctly normalised form factor: $F(0)=Z$. As noted in Ref.~\cite{Chen:1999tn} and treated in more detail in Refs.~\cite{Phillips:1999hh,Beane:2000fi}, higher-order terms in the effective-range expansion contibute only to the bound-state energy (and hence $\gamma$) and the normalisation factor multiplying all the $q$-dependence in the form factor. The first term in this expression is the part of the form factor arising from the tail of the relative wave function. The final term is the contribution of short-distance physics that is not encoded in that wave function.

\section{Summary}

In this work we have applied the Wilsonian RG to EM current operators in the pionless EFT, a theory of nonrelativistic particles interacting via short-range interactions. 
A key ingredient in deriving the RG equation for the interaction current density is the requirement that the one-particle-irreducible 5-point amplitude be independent of the cutoff \cite{Kvinikhidze:2007eu}. It is also important to maintain gauge invariance for coulings of EM fields in a theory with a cutoff. This can conveniently be done using the gauging of equations method \cite{Kvinikhidze:2009be}.

We find fixed-point solutions to the RG equation, and we study perturbations around these to determine how they scale with the cutoff. As in the case of the potential \cite{Birse:1998dk}, this scaling behaviour determines the power counting for the contributions of these operators to observables.

We have studied in particular the nontrivial fixed point describing unitary limit. 
Perturbations in the potential around this point generate the effective-range expansion. The power counting for the terms in the charge density shows that these are enhanced by two orders compared to naive dimensional analysis, matching what has been found in previous work using different methods \cite{Kaplan:1998sz,Chen:1999tn,Valderrama:2014vra}. Based on this, we present general expressions for the bremsstrahlung amplitude and bound-state form factor. These include terms that can be related to the effective-range expansion by Siegert's theorem as well as short-distance contributions that are independent of the two-body interaction. 
We also show that off-shell terms in the potential make no contributions to observables such as the on-shell bremsstrahlung amplitude.

This first application of the method demonstrates that the Wilsonian RG can be applied not only to potentials but also to other operators in nonrelativistic EFTs. We have focussed here on the case of two particles in an $S$ wave with total spin zero but it should be straightforward to extend the approach to magnetic interactions and higher partial waves. Both will introduce additional, purely transverse interactions, which are not constrained by gauge invariance. Further extensions to systems with long-range interactions will involve combining the method with the distorted-wave approach of Ref.~\cite{Barford:2002je}.

\section*{Acknowledgments}

This work was supported by the UK STFC under grants ST/L005794/1 and ST/P004423/1,
and by the Georgian Shota Rustaveli National Science Foundation (Grant No.~FR17-354).
MCB is grateful to the Institute for Nuclear Theory, Seattle for its hospitality 
during the program INT-18-2a ``Fundamental Physics with Electroweak Probes of Light Nuclei'', where part of this work was completed.

\appendix

\section{Off-shell perturbations}\label{off-shell}

In the body of this paper, we have concentrated on perturbations that depend only on energy. This is because, in the expansion around the unitary fixed point, the energy-dependent ones appear at lower orders \cite{Birse:1998dk}. In addition, there is a direct correspondence between energy-dependent terms in the potenital and terms in the effective-range expansion. In contrast, the momentum-dependent perturbations affect only the off-shell behaviour of the $T$ matrix, making no contribution to observables.
In this appendix, we examine momentum-dependent terms in the interaction current.
In particular we check that the bremsstrahlung amplitude is independent of the off-shell form of the porential in the pionless EFT, as required by general principles 
\cite{Fearing:1999im}.

For simplicity, we focus on the leading off-shell term in the potential. This has the (rescaled) form
\be
\phi_2(\hat k',\hat k; \hat p)= \left(\hat k'^2+\hat k^2-2\hat p^2
+\frac{2}{3}\,\widehat V_F(\hat p)\right)\widehat V_F(\hat p),
\label{phi2-off}
\ee
and an RG eigenvalue $\nu=+2$ \cite{Birse:1998dk}. Note that purely momentum-dependent perturbations, as considered in Ref.~\cite{Kaplan:1998tg} for example, can be expressed as sums of terms like this and energy-dependent ones. The contributions to observables come from the energy-dependent terms, not the higher-order off-shell pieces, and so it is the former that govern the power counting.

The corresponding term in the interaction charge density is
\begin{align}
\phi_2^0(\hat k',\hat k;\hat q;\hat p', \hat p)
=&\;\widehat V_F^0(\hat q;\hat p',\hat p)\left[ \hat k'^2+\hat k^2-\hat p'^2-\hat p^2
+\frac{2}{3} \left(\widehat V_F(\hat p')+\widehat V_F(\hat p)\right)\right] \nn
&\;+Z\left(\widehat V_F(\hat p')+ \widehat V_F(\hat p)\right) 
+\widehat N^0(\hat q;\hat p',\hat p),
\label{resc--cur*}
\end{align}
where
\be
\widehat N^0(\hat q;\hat p',\hat p)
=-\,\frac{Z}{8\pi}\, \widehat V_F(\hat p')\int \mathrm{d}^3\hat{\bm u}
\left[\frac{\theta(1-\hat u)}{\hat p^2-(\hat{\bm u}-\hat{\bm q}/2)^2}-
\frac{\theta(1-\hat u)}{\hat p^2-\hat u^2}\right] \widehat V_F(\hat p)
+(p'\leftrightarrow p).
\ee
This has same eigenvalue, $\nu=+2$, as the term in the potential that generates it.
The corresponding contributions to the space components of the current can be found using the WTI, Eq.~(\ref{WTI-V}), as before.

A convenient ``trick" for calculating the contributions of these perturbations to observables is to express them in terms of anticommutators with the inverse propagator. This takes advantage of the fact that these are ``equation-of-motion" terms that vanish on-shell. For the potential, we can rewrite the perturbation as
\begin{align}
\phi_2&=-\left\{\widehat G_0^{-1}, \widehat V_F\right\}
+2\widehat V_F\,\theta\,\widehat V_F\nn
&=-\left\{\widehat G_0^{-1}, \widehat V_F\right\}
+\widehat V_F\left\{\widehat G_0^{-1}, \widehat{G_0\theta} \right\} \widehat V_F,
\end{align}
where we have used $\int \hat u^2\,\mathrm{d}\hat u\, \theta(1-\hat u)=1/3$.

The perturbation in the charge density can be obtained by gauging this expression, being careful to maintain the ordering of the operators in the anticommutators. This gives
\be
\phi_2^0=-\left\{\widehat G_0^{-1}, \widehat V_F^0\right\}
+2 \widehat V_F^0\,\theta\, \widehat V_F+2 \widehat V_F\,\theta\, \widehat V_F^0
+\left\{\Gamma^0,\widehat V_F^0\right\}
+\widehat V_F \left\{\widehat G_0^{-1},\widehat{G_0\theta}\right\}^0 \widehat V_F.
\ee
The first four terms can be easily recognised as forming the first two terms of Eq.~(\ref{resc--cur*}). With some effort, the final term of Eq.~(\ref{resc--cur*}) can be expressed in a form that shows that it is $\widehat N^0$.
This term can be expanded using
\be
\left\{\widehat G_0^{-1},\widehat{G_0\theta}\right\}^0
=-\left\{\Gamma^0, \widehat{G_0\theta}\} \right\} 
+\left\{\widehat G_0^{-1},(\widehat{G_0\theta})^0\right\}
\ee
and $\widehat V_F^{-1}=\la\widehat{G_0\theta}\ra$, to get $\phi_2^0$ in the form
\be
\phi_2^0=-\left\{\widehat G_0^{-1}, \widehat V_F^0\right\}
+2 \widehat V_F^0\,\theta\, \widehat V_F+2 \widehat V_F\,\theta\, \widehat V_F^0
+\widehat V_F \left\{\widehat G_0^{-1},(\widehat{G_0\theta})^0\right\} \widehat V_F.
\label{phi0-off}
\ee

The contribution of $\phi_2$ to the rescaled $T$ matrix is
\begin{align}
\delta \widehat T&=C_2\left(1+ \widehat T_F\, \widehat{G_0^+\theta}\right)\phi_2
\left(1+ \widehat{G_0^+\theta}\, \widehat T_F\right)    \nn
&=-C_2\left\{\widehat G_0^{-1}, \widehat T_F\right\}
-C_2\widehat T_F\left\{\widehat G_0^{-1},\widehat{G_0^+\theta}
-\widehat{G_0\theta} \right\} T_F. 
\end{align}
Since the factor of $G_0^{-1}$ in the final term cancels their poles, the two propagators cancel there. This leaves only terms with external factors of $G_0^{-1}$, which vanish when they act on on-shell states:
\be
\delta \widehat T=-C_2\left\{\widehat G_0^{-1}, \widehat T_F\right\}.
 \label{delT-off}
\ee

Similarly, the contribution of this perturbation to the 5-point amplitude is
\begin{align}
\delta \widehat T^0=&\;\delta \widehat T \,\widehat{G_0^+\theta}\, V_F^0 
\left(1+ \widehat{G_0^+\theta}\, \widehat T_F\right)
+\left(1+ \widehat T_F\, \widehat{G_0^+\theta}\right)V_F^0 \,
\widehat{G_0^+\theta}\,\delta \widehat T\nn
&\;+\delta \widehat T\, (\widehat{G_0^+\theta})^0\, \widehat T_F
+ \widehat T_F\, (\widehat{G_0^+\theta})^0\,\delta \widehat T\nn
&\;+C_2\left(1+ \widehat T_F\, \widehat{G_0^+\theta}\right)\phi_2^0
\left(1+ \widehat{G_0^+\theta}\, \widehat T_F\right) .
\end{align}
Inserting the forms of $\phi_0$ and $\delta\widehat T$ from Eqs.~(\ref{phi0-off},\ref{delT-off}), this reduces to
\be
\delta\widehat T^0=-C_2\left\{\widehat G_0^{-1}, \widehat T_F^0\right\}
-C_2\,T_F\left\{\widehat G_0^{-1},(\widehat{G_0^+\theta})^0-(\widehat{G_0\theta})^0
\right\}T_F.
\ee

To simplify this, we use the symmetric prescription for the gauged regulated propagators in Eq.~(\ref{GI-Lam}) to get
\be
\left\{\widehat G_0^{-1},(\widehat{G_0\theta})^0\right\} 
=\left\{\Gamma^0, \widehat{G_0\theta}\right\}
+\frac{1}{2}\left\{[\widehat\theta,\Gamma^0],\widehat G_0\right\}.
\ee
This applies for both boundary conditions and so we can write
\be
\left\{\widehat G_0^{-1},(\widehat{G_0^+\theta})^0-(\widehat{G_0\theta})^0\right\} 
=\left\{\Gamma^0, \widehat{G_0^+\theta}-\widehat{G_0\theta}\right\}
+\frac{1}{2}\left\{[\widehat\theta,\Gamma^0],\widehat G_0^+-\widehat G_0\right\}.
\ee
Inserted between (momentum-independent) factors of $T_F$ to get the contribution to $T^\mu$, both terms contain loop integrals over $G^+_0\theta-G_0\theta$. Since the two propagators differ only by the contribution of the pole, these integrals just pick up the residue at that pole. For the integral in the first term, the LS equation can be used to show that it is just $-\widehat T_F^{-1}$. In the second term, we need to be sure that the pole is in the momentum range covered by both $\theta$ functions in the commutator $[\widehat\theta,\Gamma^0]$ (or, in other words, the momentum transfer from the EM field does not take the system outside the cutoff). Provided it is, we pick up the same residue from both pieces of the commutator, and these cancel.

Putting all these pieces together, we get the 5-point amplitude in the form
\be
\delta\widehat T^0=-C_2\left\{\widehat G_0^{-1}, \widehat T_F^0\right\}
+C_2 \left\{\Gamma^\mu,\widehat T_F\right\}.
\label{delT0-off}
\ee
Substituting this and the $T$ matrix, Eq.~(\ref{delT-off}), into Eq.~(\ref{A-brems}) gives the contribution from this perturbation to the full bremsstrahlung amplitude. 
In that amplitude, the off-shell factors of $G_0^{-1}$ in $\delta T$ cancel the propagators on the external legs. The final terms of Eq.~(\ref{A-brems}) give  $-C_2 \{\Gamma^\mu,\widehat T_F\}$, which cancels against the second term of Eq.~(\ref{delT0-off}). This leaves just
\be
\delta A^0=-C_2\left\{\widehat G_0^{-1}, \widehat T_F^0\right\},
\ee
which vanishes between on-shell states. Other off-shell perturbations with higher powes of momenta \cite{Birse:1998dk} can be handled similarly.
As expected from general principles \cite{Fearing:1999im}, we see that the off-shell behaviour of the potential has no observable consequences. 

Finally , there can be independent off-shell contributions to the charge density. These are momentum-dependent solutions to the homogeneous version of the RG equation, Eq.~(\ref{eqV0-expl}). Again, for simplicity, we show here the explicit form of only the leading ones of these. These contain terms of second order in the off-shell momenta, $k$ and $k'$, and have the form
\be
\phi^0_{2\,lmn}(\hat k',\hat k;\hat q;\hat p',\hat p)
=\left[\left(\hat k'^2-\hat p'^2+\frac{1}{3}\,\widehat V_F(\hat p')\right) 
\widehat V_F(\hat p)+\left(\hat k',\hat p'\leftrightarrow \hat k,\hat p\right)
\right]\hat q^{2l+2}\,\hat p'^{2m}\,\hat p^{2n}.
\ee
As in the momentum-independent perturbations, Eq.~(\ref{phi0lmn}), the extra factor  of $\hat q^2$ ensures that the corresponding current density is regular as $\hat q\rightarrow 0$. Their RG eigenvalues are $\nu=2(l+m+n)+4$ and so, as in the case of the potential, the off-shell perturbations are less relevant than similar on-shell ones. The momentum-dependent factors have a similar structure to those in the off-shell terms in the potential Eq.~(\ref{phi2-off}) and, in a similar manner, they do not contribute to observables.

\end{document}